
%
%
%
%
%
%
%
%
%
%
%
%
%
%
%
%
%
%
%
\nonstopmode
\catcode`\@=11 
%
%
%
\font\seventeenrm=cmr17

\font\twelverm=cmr12
\font\ninerm=cmr9
\font\sixrm=cmr6

\font\seventeenbf=cmbx12 at 17pt
\font\fourteenbf=cmbx12 at 14pt
\font\twelvebf=cmbx12
\font\ninebf=cmbx9
\font\sixbf=cmbx6

\font\seventeeni=cmmi12 at 17pt             \skewchar\seventeeni='177
\font\fourteeni=cmmi12 at 14pt              \skewchar\fourteeni='177
\font\twelvei=cmmi12                        \skewchar\twelvei='177
\font\ninei=cmmi9                           \skewchar\ninei='177
\font\sixi=cmmi6                            \skewchar\sixi='177

\font\seventeensy=cmsy10 scaled\magstep3    \skewchar\seventeensy='60
\font\fourteensy=cmsy10 scaled\magstep2     \skewchar\fourteensy='60
\font\twelvesy=cmsy10 at 12pt               \skewchar\twelvesy='60
\font\ninesy=cmsy9                          \skewchar\ninesy='60
\font\sixsy=cmsy6                           \skewchar\sixsy='60

\font\seventeenex=cmex10 scaled\magstep3
\font\fourteenex=cmex10 scaled\magstep2
\font\twelveex=cmex10 at 12pt

\font\ninex=cmex10 at 9pt
\font\sevenex=cmex10 at 9pt
\font\sixex=cmex10 at 6pt
\font\fivex=cmex10 at 5pt

\font\seventeensl=cmsl10 scaled\magstep3
\font\fourteensl=cmsl10 scaled\magstep2
\font\twelvesl=cmsl10 scaled\magstep1
\font\ninesl=cmsl10 at 9pt
\font\sevensl=cmsl10 at 7pt
\font\sixsl=cmsl10 at 6pt
\font\fivesl=cmsl10 at 5pt

\font\seventeenit=cmti12 scaled\magstep2
\font\fourteenit=cmti12 scaled\magstep1
\font\twelveit=cmti12

\font\seventeentt=cmtt12 scaled\magstep2
\font\fourteentt=cmtt12 scaled\magstep1
\font\twelvett=cmtt12

\font\seventeencp=cmcsc10 scaled\magstep3
\font\fourteencp=cmcsc10 scaled\magstep2
\font\twelvecp=cmcsc10 scaled\magstep1
\font\tencp=cmcsc10

\newfam\cpfam

\font\seventeenss=cmss17
\font\fourteenss=cmss12 at 14pt
\font\twelvess=cmss12
\font\tenss=cmss10
\font\niness=cmss9

\font\sevenss=cmss8 at 7pt
\font\sixss=cmss8 at 6pt
\font\fivess=cmss8 at 5pt
\newfam\ssfam
\newdimen\b@gheight             \b@gheight=12pt
\newcount\f@ntkey               \f@ntkey=0
\def\f@m{\afterassignment\samef@nt\f@ntkey=}
\def\samef@nt{\fam=\f@ntkey \the\textfont\f@ntkey\relax}
\def\rm{\f@m0 }
\def\mit{\f@m1 }         
\def\cal{\f@m2 }
\def\it{\f@m\itfam}
\def\sl{\f@m\slfam}
\def\bf{\f@m\bffam}
\def\tt{\f@m\ttfam}
\def\ssf{\f@m\ssfam}
\def\caps{\f@m\cpfam}
\def\seventeenpoint{\relax
    \textfont0=\seventeenrm          \scriptfont0=\twelverm
      \scriptscriptfont0=\ninerm
    \textfont1=\seventeeni           \scriptfont1=\twelvei
      \scriptscriptfont1=\ninei
    \textfont2=\seventeensy          \scriptfont2=\twelvesy
      \scriptscriptfont2=\ninesy
    \textfont3=\seventeenex          \scriptfont3=\twelveex
      \scriptscriptfont3=\ninex
    \textfont\itfam=\seventeenit    
    \textfont\slfam=\seventeensl    
      \scriptscriptfont\slfam=\ninesl
    \textfont\bffam=\seventeenbf     \scriptfont\bffam=\twelvebf
      \scriptscriptfont\bffam=\ninebf
    \textfont\ttfam=\seventeentt
    \textfont\cpfam=\seventeencp
    \textfont\ssfam=\seventeenss     \scriptfont\ssfam=\twelvess
      \scriptscriptfont\ssfam=\niness
    \samef@nt
    \b@gheight=17pt
    \setbox\strutbox=\hbox{\vrule height 0.85\b@gheight
                                depth 0.35\b@gheight width\z@ }}
\def\fourteenpoint{\relax
    \textfont0=\fourteencp          \scriptfont0=\tenrm
      \scriptscriptfont0=\sevenrm
    \textfont1=\fourteeni           \scriptfont1=\teni
      \scriptscriptfont1=\seveni
    \textfont2=\fourteensy          \scriptfont2=\tensy
      \scriptscriptfont2=\sevensy
    \textfont3=\fourteenex          \scriptfont3=\twelveex
      \scriptscriptfont3=\tenex
    \textfont\itfam=\fourteenit     \scriptfont\itfam=\tenit
    \textfont\slfam=\fourteensl     \scriptfont\slfam=\tensl
      \scriptscriptfont\slfam=\sevensl
    \textfont\bffam=\fourteenbf     \scriptfont\bffam=\tenbf
      \scriptscriptfont\bffam=\sevenbf
    \textfont\ttfam=\fourteentt
    \textfont\cpfam=\fourteencp
    \textfont\ssfam=\fourteenss     \scriptfont\ssfam=\tenss
      \scriptscriptfont\ssfam=\sevenss
    \samef@nt
    \b@gheight=14pt
    \setbox\strutbox=\hbox{\vrule height 0.85\b@gheight
                                depth 0.35\b@gheight width\z@ }}
\def\twelvepoint{\relax
    \textfont0=\twelverm          \scriptfont0=\ninerm
      \scriptscriptfont0=\sixrm
    \textfont1=\twelvei           \scriptfont1=\ninei
      \scriptscriptfont1=\sixi
    \textfont2=\twelvesy           \scriptfont2=\ninesy
      \scriptscriptfont2=\sixsy
    \textfont3=\twelveex          \scriptfont3=\ninex
      \scriptscriptfont3=\sixex
    \textfont\itfam=\twelveit    
    \textfont\slfam=\twelvesl    
      \scriptscriptfont\slfam=\sixsl
    \textfont\bffam=\twelvebf     \scriptfont\bffam=\ninebf
      \scriptscriptfont\bffam=\sixbf
    \textfont\ttfam=\twelvett
    \textfont\cpfam=\twelvecp
    \textfont\ssfam=\twelvess     \scriptfont\ssfam=\niness
      \scriptscriptfont\ssfam=\sixss
    \samef@nt
    \b@gheight=12pt
    \setbox\strutbox=\hbox{\vrule height 0.85\b@gheight
                                depth 0.35\b@gheight width\z@ }}
\def\tenpoint{\relax
    \textfont0=\tenrm          \scriptfont0=\sevenrm
      \scriptscriptfont0=\fiverm
    \textfont1=\teni           \scriptfont1=\seveni
      \scriptscriptfont1=\fivei
    \textfont2=\tensy          \scriptfont2=\sevensy
      \scriptscriptfont2=\fivesy
    \textfont3=\tenex          \scriptfont3=\sevenex
      \scriptscriptfont3=\fivex
    \textfont\itfam=\tenit     \scriptfont\itfam=\seveni
    \textfont\slfam=\tensl     \scriptfont\slfam=\sevensl
      \scriptscriptfont\slfam=\fivesl
    \textfont\bffam=\tenbf     \scriptfont\bffam=\sevenbf
      \scriptscriptfont\bffam=\fivebf
    \textfont\ttfam=\tentt
    \textfont\cpfam=\tencp
    \textfont\ssfam=\tenss     \scriptfont\ssfam=\sevenss
      \scriptscriptfont\ssfam=\fivess
    \samef@nt
    \b@gheight=10pt
    \setbox\strutbox=\hbox{\vrule height 0.85\b@gheight
                                depth 0.35\b@gheight width\z@ }}
%
%
%
\normalbaselineskip = 15pt plus 0.2pt minus 0.1pt 
\normallineskip = 1.5pt plus 0.1pt minus 0.1pt
\normallineskiplimit = 1.5pt
\newskip\normaldisplayskip
\normaldisplayskip = 15pt plus 5pt minus 10pt 
\newskip\normaldispshortskip
\normaldispshortskip = 6pt plus 5pt
\newskip\normalparskip
\normalparskip = 6pt plus 2pt minus 1pt
\newskip\skipregister
\skipregister = 5pt plus 2pt minus 1.5pt
\newif\ifsingl@    \newif\ifdoubl@
\newif\iftwelv@    \twelv@true
\def\singlespace{\singl@true\doubl@false\spaces@t}
\def\doublespace{\singl@false\doubl@true\spaces@t}
\def\normalspace{\singl@false\doubl@false\spaces@t}
\def\Tenpoint{\tenpoint\twelv@false\spaces@t}
\def\Twelvepoint{\twelvepoint\twelv@true\spaces@t}
\def\spaces@t{\relax
      \iftwelv@ \ifsingl@\subspaces@t3:4;\else\subspaces@t1:1;\fi
       \else \ifsingl@\subspaces@t3:5;\else\subspaces@t4:5;\fi \fi
      \ifdoubl@ \multiply\baselineskip by 5
         \divide\baselineskip by 4 \fi }
\def\subspaces@t#1:#2;{
      \baselineskip = \normalbaselineskip
      \multiply\baselineskip by #1 \divide\baselineskip by #2
      \lineskip = \normallineskip
      \multiply\lineskip by #1 \divide\lineskip by #2
      \lineskiplimit = \normallineskiplimit
      \multiply\lineskiplimit by #1 \divide\lineskiplimit by #2
      \parskip = \normalparskip
      \multiply\parskip by #1 \divide\parskip by #2
      \abovedisplayskip = \normaldisplayskip
      \multiply\abovedisplayskip by #1 \divide\abovedisplayskip by #2
      \belowdisplayskip = \abovedisplayskip
      \abovedisplayshortskip = \normaldispshortskip
      \multiply\abovedisplayshortskip by #1
        \divide\abovedisplayshortskip by #2
      \belowdisplayshortskip = \abovedisplayshortskip
      \advance\belowdisplayshortskip by \belowdisplayskip
      \divide\belowdisplayshortskip by 2
      \smallskipamount = \skipregister
      \multiply\smallskipamount by #1 \divide\smallskipamount by #2
      \medskipamount = \smallskipamount \multiply\medskipamount by 2
      \bigskipamount = \smallskipamount \multiply\bigskipamount by 4 }
\def\normalbaselines{ \baselineskip=\normalbaselineskip
   \lineskip=\normallineskip \lineskiplimit=\normallineskip
   \iftwelv@\else \multiply\baselineskip by 4 \divide\baselineskip by 5
     \multiply\lineskiplimit by 4 \divide\lineskiplimit by 5
     \multiply\lineskip by 4 \divide\lineskip by 5 \fi }
\Twelvepoint  
%
\interlinepenalty=50
\interfootnotelinepenalty=5000
\predisplaypenalty=9000
\postdisplaypenalty=500
\hfuzz=1pt
\vfuzz=0.2pt
\dimen\footins=24 truecm 
\hoffset=10.5truemm 
\voffset=-8.5 truemm 
%
%
%
%
%
%
\def\footnote#1{\edef\@sf{\spacefactor\the\spacefactor}#1\@sf
      \insert\footins\bgroup\singl@true\doubl@false\Tenpoint
      \interlinepenalty=\interfootnotelinepenalty \let\par=\endgraf
        \leftskip=\z@skip \rightskip=\z@skip
        \splittopskip=10pt plus 1pt minus 1pt \floatingpenalty=20000
        \smallskip\item{#1}\bgroup\strut\aftergroup\@foot\let\next}
\skip\footins=\bigskipamount 
\dimen\footins=24truecm 
\newcount\fnotenumber
\def\clearfnotenumber{\fnotenumber=0}
\def\fnote{\advance\fnotenumber by1 \footnote{$^{\the\fnotenumber}$}}
\clearfnotenumber
%
%
\newcount\secnumber
\newcount\appnumber
\newif\ifs@c 
\newif\ifs@cd 
\s@cdtrue 
\def\unsectioned{\s@cdfalse\let\section=\subsection}
\def\clearappnumber{\appnumber=64}
\def\clearsecnumber{\secnumber=0}
\newskip\sectionskip         \sectionskip=\medskipamount
\newskip\headskip            \headskip=8pt plus 3pt minus 3pt
\newdimen\sectionminspace    \sectionminspace=10pc
\newdimen\referenceminspace  \referenceminspace=25pc
\def\Titlestyle#1{\par\begingroup \interlinepenalty=9999
     \leftskip=0.02\hsize plus 0.23\hsize minus 0.02\hsize
     \rightskip=\leftskip \parfillskip=0pt
     \advance\baselineskip by 0.5\baselineskip
     \hyphenpenalty=9000 \exhyphenpenalty=9000
     \tolerance=9999 \pretolerance=9000
     \spaceskip=0.333em \xspaceskip=0.5em
     \seventeenpoint
  \noindent #1\par\endgroup }
\def\titlestyle#1{\par\begingroup \interlinepenalty=9999
     \leftskip=0.02\hsize plus 0.23\hsize minus 0.02\hsize
     \rightskip=\leftskip \parfillskip=0pt
     \hyphenpenalty=9000 \exhyphenpenalty=9000
     \tolerance=9999 \pretolerance=9000
     \spaceskip=0.333em \xspaceskip=0.5em
     \fourteenpoint
   \noindent #1\par\endgroup }
%
\def\spacecheck#1{\dimen@=\pagegoal\advance\dimen@ by -\pagetotal
   \ifdim\dimen@<#1 \ifdim\dimen@>0pt \vfil\break \fi\fi}
\def\section#1{\cleareqnumber \s@ctrue \global\advance\secnumber by1
   \par \ifnum\the\lastpenalty=30000\else
   \penalty-200\vskip\sectionskip \spacecheck\sectionminspace\fi
   \noindent {\caps\enspace\S\the\secnumber\quad #1}\par
   \nobreak\vskip\headskip \penalty 30000 }
\def\subsection#1{\par
   \ifnum\the\lastpenalty=30000\else \penalty-100\smallskip
   \spacecheck\sectionminspace\fi
   \noindent\undertext{#1}\enspace \vadjust{\penalty5000}}

\def\undertext#1{\vtop{\hbox{#1}\kern 1pt \hrule}}
\def\subsubsection#1{\par
   \ifnum\the\lastpenalty=30000\else \penalty-100\smallskip \fi
   \noindent\hbox{#1}\enspace \vadjust{\penalty5000}}

\def\appendix#1{\cleareqnumber \s@cfalse \global\advance\appnumber by1
   \par \ifnum\the\lastpenalty=30000\else
   \penalty-200\vskip\sectionskip \spacecheck\sectionminspace\fi
   \noindent {\caps\enspace Appendix \char\the\appnumber\quad #1}\par
   \nobreak\vskip\headskip \penalty 30000 }
\clearsecnumber
\clearappnumber
%
%
\def\ack{\par\penalty-100\medskip \spacecheck\sectionminspace
   \line{\iftwelv@\fourteencp\else\twelvecp\fi\hfil ACKNOWLEDGEMENTS\hfil}%
\nobreak\vskip\headskip }
\def\refs{\begingroup \par\penalty-100\medskip \spacecheck\sectionminspace
   \line{\iftwelv@\fourteencp\else\twelvecp\fi\hfil REFERENCES\hfil}%
\nobreak\vskip\headskip \frenchspacing }
\def\endrefs{\par\endgroup}
%
\newcount\refnumber
\def\clearrefnumber{\refnumber=0}  \clearrefnumber
\newwrite\R@fs                              
\immediate\openout\R@fs=\jobname.references 
\def\closerefs{\immediate\closeout\R@fs} 
\def\refsout{\closerefs\refs
\catcode`\@=11                          
\input\jobname.references               
\catcode`\@=12			        
\endrefs}
\def\refitem#1{\item{{\bf #1}}}
\def\ifundefined#1{\expandafter\ifx\csname#1\endcsname\relax}
%
%
\def\[#1]{\ifundefined{#1R@FNO}%
\global\advance\refnumber by1%
\expandafter\xdef\csname#1R@FNO\endcsname{[\the\refnumber]}%
\immediate\write\R@fs{\noexpand\refitem{\csname#1R@FNO\endcsname}%
\noexpand\csname#1R@F\endcsname}\fi{\bf \csname#1R@FNO\endcsname}}
\def\refdef[#1]#2{\expandafter\gdef\csname#1R@F\endcsname{{#2}}}
%
%
%
%
%
%
\newcount\eqnumber
\def\cleareqnumber{\eqnumber=0}
\newif\ifal@gn \al@gnfalse  
\def\veqnalign#1{\al@gntrue \vbox{\eqalignno{#1}} \al@gnfalse}
\def\eqnalign#1{\al@gntrue \eqalignno{#1} \al@gnfalse}
\def\(#1){\relax%
\ifundefined{#1@Q}
 \global\advance\eqnumber by1
 \ifs@cd
  \ifs@c
   \expandafter\xdef\csname#1@Q\endcsname{{%
\noexpand\rm(\the\secnumber .\the\eqnumber)}}
  \else
   \expandafter\xdef\csname#1@Q\endcsname{{%
\noexpand\rm(\char\the\appnumber .\the\eqnumber)}}
  \fi
 \else
  \expandafter\xdef\csname#1@Q\endcsname{{\noexpand\rm(\the\eqnumber)}}
 \fi
 \ifal@gn
    & \csname#1@Q\endcsname
 \else
    \eqno \csname#1@Q\endcsname
 \fi
\else%
\csname#1@Q\endcsname\fi\global\let\@Q=\relax}
%
%
\newif\iffrontpage \frontpagefalse
\headline={\hfil}
\footline={\iffrontpage\hfil\else \hss\twelverm
-- \folio\ --\hss \fi }
\def\monthname{\relax\ifcase\month 0/\or January\or February\or
   March\or April\or May\or June\or July\or August\or September\or
   October\or November\or December\else\number\month/\fi}
\hsize=14 truecm
\vsize=22 truecm
\skip\footins=\bigskipamount
\normalspace
%
%
%
\newskip\frontpageskip
\newif\ifp@bblock \p@bblocktrue
\newif\ifm@nth \m@nthtrue
\newtoks\pubnum
\newtoks\pubtype
\newtoks\m@nthn@me
\newcount\Ye@r
\advance\Ye@r by \year
\advance\Ye@r by -1900
\def\Year#1{\Ye@r=#1}
\def\Month#1{\m@nthfalse \m@nthn@me={#1}}
\def\m@nthname{\ifm@nth\monthname\else\the\m@nthn@me\fi}
\def\titlepage{\global\frontpagetrue\hrule height\z@ \relax
               \ifp@bblock\pubblock\fi\relax }
\def\endtitlepage{\vfil\break
                  \frontpagefalse} 
\def\bonntitlepage{\global\frontpagetrue\hrule height\z@ \relax
               \ifp@bblock\pubblock\fi\relax }
\frontpageskip=12pt plus .5fil minus 2pt
\pubtype={\iftwelv@\twelvesl\else\tensl\fi\ (Preliminary Version)}
\pubnum={?}
\def\nopubblock{\p@bblockfalse}
\def\pubblock{\line{\hfil\iftwelv@\twelverm\else\tenrm\fi%
BONN--HE--\number\Ye@r--\the\pubnum\the\pubtype}
              \line{\hfil\iftwelv@\twelverm\else\tenrm\fi%
\m@nthname\ \number\year}}
\def\title#1{\vskip\frontpageskip\Titlestyle{\caps #1}\vskip3\headskip}
\def\author#1{\vskip.5\frontpageskip\titlestyle{\caps #1}\nobreak}

\def\authors{\vskip\frontpageskip\noindent}
\def\address#1{\par\kern 5pt\titlestyle{
\it #1}}
\def\andaddress{\par\kern 5pt \centerline{\sl and} \address}
\def\addresses{\vskip\frontpageskip\noindent\interlinepenalty=9999}
\def\abstract#1{\par\dimen@=\prevdepth \hrule height\z@ \prevdepth=\dimen@
   \vskip\frontpageskip\spacecheck\sectionminspace
   \centerline{\iftwelv@\fourteencp\else\twelvecp\fi ABSTRACT}\vskip\headskip
   {\noindent #1}}
%

%
%
%
\def\leaderfill{\leaders\hbox to 1em{\hss.\hss}\hfill}
\def\boxit#1{\vcenter{\hrule\hbox{\vrule\kern8pt
      \vbox{\kern8pt#1\kern8pt}\kern8pt\vrule}\hrule}}

%
%
%
\def\ref#1{{\bf [#1]}}
\def\th{{\rm th}}
\def\nl{\hfil\break}
%
%
%
%
%
\newif\ifm@thstyle \m@thstylefalse
\def\mathstyle{\m@thstyletrue}
\def\proclaim#1#2\par{\smallbreak\begingroup
\advance\baselineskip by -0.25\baselineskip%
\advance\belowdisplayskip by -0.35\belowdisplayskip%
\advance\abovedisplayskip by -0.35\abovedisplayskip%
    \noindent{\caps#1.\enspace}{#2}\par\endgroup%
\smallbreak}
\def\m@kem@th<#1>#2#3{%
\ifm@thstyle \global\advance\eqnumber by1
 \ifs@cd
  \ifs@c
   \expandafter\xdef\csname#1\endcsname{{%
\noexpand #2\ \the\secnumber .\the\eqnumber}}
  \else
   \expandafter\xdef\csname#1\endcsname{{%
\noexpand #2\ \char\the\appnumber .\the\eqnumber}}
  \fi
 \else
  \expandafter\xdef\csname#1\endcsname{{\noexpand #2\ \the\eqnumber}}
 \fi
 \proclaim{\csname#1\endcsname}{#3}
\else
 \proclaim{#2}{#3}
\fi}
%
%
%
%
%
%
\def\Thm<#1>#2{\m@kem@th<#1M@TH>{Theorem}{\sl#2}}
\def\Prop<#1>#2{\m@kem@th<#1M@TH>{Proposition}{\sl#2}}
\def\Def<#1>#2{\m@kem@th<#1M@TH>{Definition}{\rm#2}}
\def\Lem<#1>#2{\m@kem@th<#1M@TH>{Lemma}{\sl#2}}
\def\Cor<#1>#2{\m@kem@th<#1M@TH>{Corollary}{\sl#2}}
\def\Conj<#1>#2{\m@kem@th<#1M@TH>{Conjecture}{\sl#2}}
\def\Rmk<#1>#2{\m@kem@th<#1M@TH>{Remark}{\rm#2}}
\def\Exm<#1>#2{\m@kem@th<#1M@TH>{Example}{\rm#2}}
\def\Qry<#1>#2{\m@kem@th<#1M@TH>{Query}{\it#2}}
%
%

\def\<#1>{\csname#1M@TH\endcsname}
%
%

\def\implies{\Rightarrow}
\def\lapprox{\hbox{\lower3pt\hbox{$\buildrel<\over\sim$}}}
\def\gapprox{\hbox{\lower3pt\hbox{$\buildrel<\over\sim$}}}
\def\quotient#1#2{#1/\lower0pt\hbox{${#2}$}}

%
%
\def\to{\rightarrow}
%

\def\mapdown#1{\Big\downarrow
  \rlap{$\vcenter{\hbox{$\scriptstyle#1$}}$}}
\def\mapup#1{\Big\uparrow
  \rlap{$\vcenter{\hbox{$\scriptstyle#1$}}$}}

\def\commdiag#1{
\def\normalbaselines{\baselineskip20pt \lineskip3pt \lineskiplimit3pt }
\matrix{#1}} 
%
%
\def\reals{\mathord{\bf R}} 
\def\integ{\mathord{\bf Z}} 
\def\nats{\mathord{\bf N}} 
%
%
\def\Tr{\mathop{\rm Tr}}
\def\underrightarrow#1{\vtop{\ialign{##\crcr
      $\hfil\displaystyle{#1}\hfil$\crcr
      \noalign{\kern-\p@\nointerlineskip}
      \rightarrowfill\crcr}}} 
\def\underleftarrow#1{\vtop{\ialign{##\crcr
      $\hfil\displaystyle{#1}\hfil$\crcr
      \noalign{\kern-\p@\nointerlineskip}
      \leftarrowfill\crcr}}}  

%
%
\def\comm#1#2{\left[#1\, ,\,#2\right]}
%
%
\def\pder#1#2{{{\partial #1}\over{\partial #2}}}
%
%
%
%
%
%
\newdimen\unit
\newdimen\redunit
%
%
\def\p@int#1:#2 #3 {\rlap{\kern#2\unit
     \raise#3\unit\hbox{#1}}}
%
%
\def\th@r{\vrule height0\unit depth.1\unit width1\unit}
\def\bh@r{\vrule height.1\unit depth0\unit width1\unit}
\def\lv@r{\vrule height1\unit depth0\unit width.1\unit}
\def\rv@r{\vrule height1\unit depth0\unit width.1\unit}
%
%
\def\t@ble@u{\hbox{\p@int\bh@r:0 0
                   \p@int\lv@r:0 0
                   \p@int\rv@r:.9 0
                   \p@int\th@r:0 1
                   }
             }
%
%
\def\t@bleau#1#2{\rlap{\kern#1\redunit
     \raise#2\redunit\t@ble@u}}
%
%
\newcount\n
\newcount\m
\def\makecol#1#2#3{\n=0 \m=#3
  \loop\ifnum\n<#1{}\advance\m by -1 \t@bleau{#2}{\number\m}\advance\n by 1
\repeat}
%
%
\def\makerow#1#2#3{\n=0 \m=#3
 \loop\ifnum\n<#1{}\advance\m by 1 \t@bleau{\number\m}{#2}\advance\n by 1
\repeat}
%
%
\def\checkunits{\ifinner \unit=6pt \else \unit=8pt \fi
                \redunit=0.9\unit } 
\def\ytsym#1{\checkunits\kern-.5\unit
  \vcenter{\hbox{\makerow{#1}{0}{0}\kern#1\unit}}\kern.5em} 
\def\ytant#1{\checkunits\kern.5em
  \vcenter{\hbox{\makecol{#1}{0}{0}\kern1\unit}}\kern.5em} 
\def\ytwo#1#2{\checkunits
  \vcenter{\hbox{\makecol{#1}{0}{0}\makecol{#2}{1}{0}\kern2\unit}}
                  \ } 
\def\ythree#1#2#3{\checkunits
  \vcenter{\hbox{\makecol{#1}{0}{0}\makecol{#2}{1}{0}\makecol{#3}{2}{0}%
\kern3\unit}}
                  \ } 
%
%
%

\def\NPB#1#2#3{{\sl Nucl. Phys.} {\bf B#1} (#2) #3}

\def\CMP#1#2#3{{\sl Comm. Math. Phys.} {\bf #1} (#2) #3}

\def\PLB#1#2#3{{\sl Phys. Lett.} {\bf #1B} (#2) #3}
\def\JMP#1#2#3{{\sl J. Math. Phys.} {\bf #1} (#2) #3}

\def\RMS#1#2#3{{\sl Russian Math Surveys} {\bf #1} (#2) #3}

\def\FAaIA#1#2#3{{\sl Functional Analysis and Its Application} {\bf #1} (#2)
#3}

\def\Invm#1#2#3{{\sl Invent. math.} {\bf #1} (#2) #3}
\def\LMP#1#2#3{{\sl Letters in Math. Phys.} {\bf #1} (#2) #3}
\def\IJMPA#1#2#3{{\sl Int. J. Mod. Phys.} {\bf A#1} (#2) #3}

\def\PJAS#1#2#3{{\sl Proc. Jpn. Acad. Sci.} {\bf #1} (#2) #3}
\def\JPSJ#1#2#3{{\sl J. Phys. Soc. Jpn.} {\bf #1} (#2) #3}

\def\AdiM#1#2#3{{\sl Annali di Matematica} {\bf #1} (#2) #3}
\catcode`\@=12 
%
%
%
%
\def\W{\mathord{\ssf W}}
\def\w{\mathord{\ssf w}}
\def\diffs#1{\mathord{\ssf diff}(S^{#1})}
\def\GD{\mathord{\ssf GD}}
\def\gd{\mathord{\ssf gd}}

\def\Winfty{\W_\infty}
\def\Wkpq{\W_{\rm KP}^{(q)}}
\let\pb=\anticomm
\def\winfty{\w_\infty}
\let\d=\partial
\def\kpder#1#2#3{{{\partial^{#1} #2}\over{\partial #3^{#1}}}}
\def\ad{\mathop{\rm ad}}

\def\comb[#1/#2]{\left[{#1\atop#2}\right]}
\def\fr#1/#2{\ifinner{{\scriptstyle#1}\over{\scriptstyle#2}}\else%
                      {{#1}\over{#2}}\fi}
\def\mod{\mathop{\rm mod}}

\def\utilde{\widetilde{u}}

\def\longequal{\buildrel\hbox to\wd1{\hrulefill}\over{\hbox to
\wd1{\hrulefill}}}
\def\rflecha#1{\buildrel\hbox{\ninerm #1}\over{\hbox
to\wd1{\rightarrowfill}}}
\def\lflecha#1{\buildrel\hbox{\ninerm #1}\over{\hbox
to\wd1{\leftarrowfill}}}
\def\dlb#1#2{\lbrack\!\lbrack#1,#2\rbrack\!\rbrack}

\def\Wnew{\W_\infty^\#}
\let\isom=\cong
%
%
\refdef[GerNev]{J.-L.~Gervais and A.~Neveu, \NPB{209}{1982}{125}.}
\refdef[Magri]{F.~Magri, \JMP{19}{1978}{1156}.}
\refdef[WReview]{P.~Bouwknegt and K.~Schoutens, {\sl $\W$-Symmetry in
Conformal Field Theory},  {\it Phys. Reps.} to appear.}
\refdef[Dickey]{L.~A.~Dickey, {\sl Soliton equations and Hamiltonian
systems},  Advanced Series in Mathematical Physics Vol.12,  World
Scientific Publ.~Co..}
\refdef[FL]{V.~A.~Fateev and S.~L.~Lykyanov, \IJMPA{3}{1988}{507}.}
\refdef[Lyk]{S.~L.~Luk'yanov, \FAaIA{22}{1988}{255}.}
\refdef[winfty]{I.~Bakas, \PLB{228}{1989}{406}; \CMP{134}{1990}{487}.}
\refdef[Winfty]{C.~N.~Pope, L.~J.~Romans, and X.~Shen,
\NPB{339}{1990}{191}.}
\refdef[BKK]{I.~Bakas, B.~Khesin and E.~Kiritsis, Preprint
UCB-PTH-91/48, LBL-31303, UMD-PP-92-36.}
\refdef[Woneplusinfty]{C. N. Pope, L. J. Romans, and X. Shen,
\PLB{242}{1990}{401}.}
\refdef[PRS]{C. N. Pope, L. J. Romans, and X. Shen,
\PLB{245}{1990}{72}.}
\refdef[KK]{B.~A.~Khesin and O.~S.~Kravchenko, \FAaIA{25}{1991}{152}.}
\refdef[Feigin]{B.~Feigin, \RMS{43}{1988}{157}.}
\refdef[Watanabe]{Y.~Watanabe, \LMP{7}{1983}{99};
\AdiM{136}{1984}{77}.}
\refdef[WinftyKP]{K.~Yamagishi, \PLB{259}{1991}{436};\nl F.~Yu and
Y.-S.~Wu, \PLB{236}{1991}{220}}
\refdef[KP]{E.~Date, M.~Jimbo, M.~Kashiwara, and T.~Miwa
\PJAS{57A}{1981}{387}; \JPSJ{50}{1981}{3866}.}
\refdef[GD]{I.~M.~Gel'fand and L.~A.~Dickey, {\sl A family of
Hamiltonian structures connected with integrable nonlinear
differential equations}, Preprint 136, IPM AN SSSR, Moscow (1978).}
\refdef[AddSym]{S.~Aoyama and Y.~Kodama, \PLB{278}{1992}{56}.}
\refdef[WKP]{J.~M.~Figueroa-O'Farrill, J.~Mas, and E.~Ramos,
\PLB{266}{1991}{298}.}
\refdef[DickeyKP]{L.~A.~Dickey, {\sl Annals NY Acad.~Sci.} {\bf
491}(1987) 131.}
\refdef[WuYuKP]{F.~Yu and Y.-S.~Wu, \NPB{373}{1992}{713}.}
\refdef[Univ]{J.~M.~Figueroa-O'Farrill and E.~Ramos,
\JMP{33}{1992}{833}.}
\refdef[Class]{J.~M.~Figueroa-O'Farrill and E.~Ramos, \PLB{282}{1992}{357}
({\tt hep-th/9202040}); {\sl Classical $\W$-algebras from
dispersionless Lax hierarchies}, preprint to appear.}
\refdef[WKPq]{J.~M.~Figueroa-O'Farrill, J.~Mas, and E.~Ramos, {\sl A
One-Parameter Family of Hamiltonian Structures for the KP Hierarchy
and a Continuous Deformation of the Nonlinear $\W_{\rm KP}$ Algebra},
Preprint BONN--HE--92--20, US--FT--92/7, KUL--TF--92/20,
{\tt hep-th/9207092}.}
\refdef[Adler]{M.~Adler, \Invm{50}{1979}{403}.}
\refdef[Didier]{D.~Depireux, Preprint LAVAL PHY-21-92, {\tt
hepth@xxx/9203062}.}
\refdef[Radul]{A.~O.~Radul, in {\sl Applied methods of nonlinear
analysis and control}, pp. 149-157, Mironov, Moroz, and Tshernjatin,
eds.,  MGU 1987 (Russian).}
\refdef[Das]{A.~Das and W.-J.~Huang, \JMP{33}{1992}{2487}.}
\refdef[dFIZ]{P.~Di Francesco, C.~Itzykson, and J.-B.~Zuber,
\CMP{140}{1991}{543}.}
\unsectioned
\overfullrule=0pt
\def\pubblock{ \line{\hfil\twelverm BONN--HE--92--24}
               \line{\hfil\twelverm US--FT--8/92}
               \line{\hfil\twelverm KUL--TF--92/32}
               \line{\hfil\twelvett hep-th/9208077}
               \line{\hfil\twelverm August 1992}}
\titlepage
\title{The Topography of $\Winfty$-type Algebras}
\authors
{\caps Jos\'e~M.~Figueroa-O'Farrill$^1$\footnote{$^\flat$}{\tt
e-mail: figueroa@pib1.physik.uni-bonn.de}},
{\caps Javier~Mas$^2$\footnote{$^\natural$}{\tt e-mail:
jamas@gaes.usc.es}}, and {\caps
Eduardo~Ramos$^3$\footnote{$^\sharp$}{{\tt e-mail:
fgbda06@blekul11.bitnet} {\rm (Address after October 1992: Queen Mary
and Westfield College, UK)}}}
\addresses
$^1${\it Physikalisches Institut der Universit\"at Bonn, Nu{\ss}allee
12, W-5300 Bonn 1, GERMANY}\hfil\break\noindent
$^2${\it Departamento de F{\'\i}sica de Part{\'\i}culas Elementales,
Universidad de Santiago, E-15706 Santiago de Compostela,
SPAIN}\hfil\break\noindent
$^3${\it Instituut~voor~Theoretische~Fysica, Universiteit~Leuven,
Celest\"ynenlaan 200D, B-3001~Heverlee, BELGIUM}
\abstract{We chart out the landscape of $\Winfty$-type algebras using
$\Wkpq$---a recently discovered one-parameter deformation of $\W_{\rm
KP}$.  We relate all hitherto known $\Winfty$-type algebras to $\Wkpq$
and its reductions, contractions, and/or truncations at special values
of the parameter.}
\endtitlepage

\section{Introduction and History}

Ever since the realization \[GerNev] that the Magri bracket \[Magri]
for the KdV hierarchy could be identified with the Virasoro algebra,
the study of classical integrable systems and $\W$-algebras (see
\[Dickey] and \[WReview] respectively for up-to-date reviews on the
subjects) have been inextricably linked to each other.  For instance,
the realization that the second Gel'fand--Dickey bracket \[GD] for the
Boussinesque hierarchy could be identified with Zamolodchikov's
$\W_3$, lead Fateev and Lykyanov \[FL] to construct the
generalisations known as $\W_n$ by quantizing \[Lyk] the second
Gel'fand--Dickey bracket for the $n^\th$-order KdV hierarchy.  These
are extensions of the Virasoro algebra by primary fields of spins
$3,4,\ldots, n$.  This example is prototypical: classical realizations
of $\W$-algebras appear as hamiltonian structures of integrable
hierarchies and their quantization is made possible by quantizing free
field realizations defined at the classical level via Miura-type
transformations.

In this letter we shall focus on $\W$-algebras containing an infinite
tower of higher spin generators.  For convenience we refer to these
algebras as being of the $\Winfty$-type.  The study of these algebras
was started in \[winfty] where the algebra $\winfty$ was obtained as a
contraction $n\to\infty$ of $\W_n$ and was identified with a
subalgebra of the Lie algebra of area-preserving diffeomorphisms on
the cylinder.  Just like any other 2-manifold, the cylinder is
symplectic and area-preserving diffeomorphisms are simply
symplectomorphisms, and can thus be identified locally with the
Poisson algebra of smooth functions.  If $z,\xi$ are local Darboux
coordinates on the cylinder $S^1\times \reals$, then a basis for the
functions is given by $\{z^n \xi^s\}$, where $n\in\integ$ and
$s\in\nats$.  They obey a Lie algebra defined by the fundamental
Poisson bracket $\pb{\xi}{z}=1$.  The algebra $\winfty$ is the
subalgebra generated by those $z^n \xi^s$ with $s\geq 1$.  Being a
limit $n\to\infty$ of $\W_n$ we expect that it contains
generators of all spins $s\geq 2$.  This is indeed the case: if we let
$W^{(s)}_n$ for $s\geq 2$ and $n\in\integ$ denotes the modes of a
generator of spin $s$, and we map $W^{(s)}_n \mapsto -z^{n+1-s}
\xi^{s-1}$, the induced mode algebra is such that $W^{(2)}$ satisfies
a $\diffs1$ algebra under which $W^{(s)}$ is a primary field of spin
$s$.  This algebra, as it stands, admits a central extension only in
the Virasoro sector. The inclusion of central charges in all spin
sectors requires a modification of the algebra.

This was achieved in \[Winfty] where a deformation $\Winfty$ of
$\winfty$ was found by brute-force imposition of the Jacobi identities
within a suitable Ansatz.  With some hindsight, the construction of
$\Winfty$ appears very naturally as simply the quantization of the
functions on the cylinder.  The cylinder is symplectomorphic to the
cotangent bundle of the circle and can thus be interpreted as the
phase space of a physical system whose configuration space is a
circle.  The Poisson algebra of functions are the classical
observables, which upon quantization get mapped to the Lie algebra of
differential operators on the circle which has a unique nontrivial
central extension \[Feigin] given recently by the Khesin--Kravchenko
cocycle \[KK].  Therefore $\Winfty$ is a subalgebra of the Lie algebra
of differential operators on the circle, the mapping being $W_n^{(s)}
\mapsto -z^{n+1-s} \d^{s-1}$ for $n\in\integ$ and $s\geq 1$.  In other
words, $\Winfty$ corresponds to those differential operators on the
circle with no piece of order zero.  Adding the zeroth order piece
yields an extension of $\Winfty$ by a generator of spin 1.  This
algebra, which is clearly isomorphic to the algebra of differential
operators on the circle, is called $\W_{1+\infty}$ and was first
constructed in \[Woneplusinfty].  Its identification with the Lie
algebra of differential operators on the circle was carried out in
\[PRS] in a different context.  This identification makes manifest a
nested sequence of subalgebras of $\W_{1+\infty}$:
$$\W_{1+\infty} \supset \Winfty \supset \W_{\infty-2} \supset \cdots
\supset \W_{\infty - N} \supset \cdots~,\()$$
where $\W_{\infty-N}$ is the algebra of differential operators
of the form $\sum_{i>N} a_i\d^i$.  In terms of $\W_{1+\infty}$
they correspond to successive truncations from below: keeping only
generators of spin $>N$.  The central extension for all these algebras
can be uniformly described \[BKK] by the Khesin--Kravchenko cocycle
and we will return to this point later.

A more conclusive interpretation of $\W_{1+\infty}$ in terms of
integrable systems was found in \[WinftyKP] where it was identified
(albeit without central extension) with the first hamiltonian
structure \[Watanabe] of the KP \[KP] hierarchy.\fnote{More recently,
$\Winfty$ (also without a central extension) has been identified as
the algebra of additional symmetries of the KP hierarchy \[AddSym].}
This suggests that one should try to recover the characteristic
nonlinearity of $\W$-algebras, which is absent in $\W_{1+\infty}$ and
its truncations, by looking for the analog in the KP hierarchy of the
second Gel'fand--Dickey bracket.  This was carried out in \[WKP],
rediscovering earlier work of Dickey \[DickeyKP], where we constructed
$\W_{\rm KP}$: a nonlinear centerless deformation of $\W_{1+\infty}$.
Independently and at the same time, Wu and Yu constructed in \[WuYuKP]
a centerless nonlinear deformation of $\Winfty$ called
$\widehat{\W}_\infty$ and which can be seen as a reduction of $\W_{\rm
KP}$.  Considering the fact that the generalized KdV hierarchies are
reductions of KP, it was hoped that $\W_{\rm KP}$ (or
$\widehat{\W}_\infty$) would be universal for the $\W_n$ algebras in
the sense---made precise in \[Univ]---that all $\W_n$ algebras could
be obtained from it by reduction.  Nevertheless no $\W_n$ has ever
been constructed as a reduction of neither of these two
algebras---although, as shown in \[Class], the classical limits of
$\W_n$ can all be obtained from the classical limit of $\W_{\rm KP}$
by reduction.

This result prompted us to investigate possible deformations of
$\W_{\rm KP}$ in hopes of finding the universal $\W$-algebra.  In
\[WKPq] we found a one-parameter family of hamiltonian structures for
the KP hierarchy and, as a consequence, a one-parameter deformation
$\Wkpq$ of $\W_{\rm KP}$.  This one-parameter family relates all
hitherto known $\Winfty$-type algebras (with or without central
extensions) either as reductions, contractions, or truncations and,
moreover, it also reduces (for $q=n$) to $\W_n$.  The purpose of this
paper is to exhibit the topography of the $\Winfty$-landscape which
has emerged as a consequence of the construction of this one-parameter
family of $\W$-algebras.  We hope to convince the reader that $\Wkpq$
is a valuable and much-needed organizational tool in this field and to
this effect we summarize diagrammatically throughout the paper the
relations found between different $\Winfty$-type algebras.

We organize this paper as follows.  In then next section we briefly
review the construction of $\Wkpq$ in terms of an extension of the
Adler map to the space of generalized pseudodifferential symbols, and
its reduction to $\widehat{\W}_\infty^{(q)}$ by setting the generator
of spin 1 to zero.  We then go on to analyze the structure of $\Wkpq$
for $q=N$ a positive integer.  We show that $\W_{\rm KP}^{(N)}$ (resp.
$\widehat{\W}_\infty^{(N)}$) reduces to the Gel'fand--Dickey algebra
$\GD_N$ (resp. $\W_N$).  It also contracts to centerless versions of
$\W_{1+\infty}$, $\Winfty$, and $\W_{\infty - N}$.  We then show how
to recover the centrally extended version of these algebras as
particular contractions in which the limit $q\to N$ is approached at
the same as the generators are rescaled.  The expressions for the
limiting Poisson structures involve the Khesin--Kravchenko cocycle.
We also obtain a new nonlinear $\W$-algebra $\Wnew$ as a reduction of
$\W_{1+\infty}$ or, equivalently, as a contraction $q\to 0$ of
$\widehat{\W}_\infty^{(q)}$.  We then take a tour through the
classical limits of these algebras.  The $q$-dependence of $\Wkpq$
(resp. $\widehat{\W}_\infty^{(q)}$) becomes fictitious in this limit
and for $q\neq 0$ they yield algebras equivalent to $\w_{\rm KP}$
(resp. $\widehat{\w}_\infty$).  As contractions we obtain
$\w_{1+\infty}$, $\winfty$, as well as its further truncations
$\w_{\infty - N}$.  The classical limit of $\Wnew$ is a nonlinear
algebra inequivalent to $\widehat{\w}_\infty$.  Finally, we discuss
the covariance properties of $\Winfty$-type algebras.  Every such
algebra admits an action of the algebra of diffeomorphisms of the
circle via derivations.  For the case of $\W_N$ it was proven in
\[dFIZ] that a primary basis for the algebra exists.  It is a tacit
assumption in the literature that this continues to be the case in
$\Winfty$-type algebras---albeit at the price of introducing
nonlinearities.  However we find that this is not the case for
$\widehat{\W}_\infty$.  We conjecture that a primary basis for the
algebra exists as long as the Virasoro central charge is nonzero.

\section{The $\W_{\rm KP}^{(q)}$ Algebra}

All the $\W$-algebras that we will examine in this letter appear as
Poisson structures on differential algebras generated by a collection
$\{u_i\}$ of infinitely differentiable functions, say, on the circle.
In other words, we work with the algebra of polynomials in the
$\{u_i\}$ and their derivatives.  A $\W$-algebra generated by the
$\{u_i\}$ will then take the form of a bracket
$$\pb{u_i(z)}{u_j(w)} = J_{ij}(z)\cdot
\delta(z-w)~,\(Poissonbrackets)$$
where $J_{ij}$ are some differential operators whose coefficients are
differential polynomials in the $\{u_i\}$.  For \(Poissonbrackets) to
define a Poisson bracket we must require antisymmetry and the Jacobi
identity.  A very elegant way of generating such Poisson structures is
via the Adler map in the space of one-dimensional pseudodifferential
symbols.  We shall construct $\Wkpq$ by a slight generalization of
this procedure.

By a pseudodifferential symbol we mean a formal Laurent series in a
parameter $\xi^{-1}$ of the form $P(z,\xi) = \sum_{i=-\infty}^{\rm
finite} p_i(z)\xi^i$ whose coefficients we take to be smooth functions
on the circle.  Symbols have a commutative multiplication given by
multiplying the Laurent series; but one can also define a composition
law (denoted by $\circ$)
$$P(z,\xi)\circ Q(z,\xi)= \sum_{k\geq 0} {1\over k!}
\kpder{k}{P}{\xi} \kpder{k}{Q}{z}~,\(symbolcomp)$$
making the map
$$P(z,\xi) \mapsto \sum_i p_i(z) \d^i\(symtops)$$
from pseudodifferential symbols to pseudodifferential operators into
an algebra homomorphism.

Symbol composition is moreover a well-defined operation on arbitrary
smooth functions of $z$ and $\xi$.  For example, for $a=a(z)$,
$$\log\xi \circ a = a\log\xi - \sum_{j=1}^\infty
{(-1)^j\over j} a^{(j)} \xi^{-j}~,\(logderiv)$$
which shows that the commutator (under symbol composition) with
$\log\xi$, denoted by $\ad\log\xi$, is an outer derivation on the
algebra of pseudodifferential symbols.  Similarly, if $q$ is any
complex number, not necessarily an integer, we find
$$\xi^q \circ a = \sum_{j=0}^\infty \comb[q/j] a^{(j)}
\xi^{q-j}~,\(qLeibniz)$$
where we have introduced, for $q$ any complex number, the generalized
binomial coefficients $\comb[q/j] \equiv [q]_j/j!$, where $[q]_j
\equiv q(q-1)\cdots(q-j+1)$ is the Pochhammer symbol.  Conjugation by
$\xi^q$ is therefore an outer automorphism of the algebra of
pseudodifferential symbols, which is the integrated version of $\ad
\log\xi$.

It follows from \(qLeibniz) that (left and right) multiplication by
$\xi^q$ sends pseudodifferential symbols into symbols of the form
$\sum_{j\leq N} p_j(z) \xi^{q+j}$.  Let us denote the set of these
symbols by ${\cal S}_q$.  It is clear that ${\cal S}_q$ is a bimodule
over the algebra of pseudodifferential symbols, which for
$q\in\integ$ coincides with the algebra itself.  In fact, since ${\cal
S}_q = {\cal S}_p$ for $p \equiv q \mod \integ$, we will understand
${\cal S}_q$ from now on as implying that $q$ is reduced modulo the
integers.  Moreover, symbol composition induces a multiplication
${\cal S}_p \times {\cal S}_q \to {\cal S}_{p+q}$, where we add modulo
the integers.  We call their union ${\cal S}=\cup_q {\cal S}_q$, the
algebra of generalized pseudodifferential symbols.  Furthermore, the
algebra ${\cal S}_0$ of pseudodifferential symbols splits into the
direct sum of two subalgebras ${\cal S}_0 = {\cal R}_+ \oplus {\cal
R}_-$, corresponding to the differential and integral symbols
respectively.  We denote by $A_\pm$ the projection of $A$ onto ${\cal
R}_\pm$ along ${\cal R}_\mp$.

The construction of $\Wkpq$ now runs as follows.  Put the generators
$\{u_i\}$ into the following generalized pseudodifferential symbol:
$$L = \alpha \xi^q + \sum_{i=1}^\infty u_i \xi^{q-i} \in {\cal
S}_q~,\(qLaxOp)$$
where $\alpha$ is an inessential parameter which we introduce for
later convenience.  Now let $X=\sum_{i=1}^\infty \xi^{i-q-1}\circ x_i
\in {\cal S}_{-q}$ and define the following generalization of the
Adler map \[Adler]:
$$\veqnalign{J^{(q)}(X) &= {1\over\alpha}\left[ (L\circ X)_+\circ L - L\circ
(X\circ L)_+\right]\cr
 &= {1\over\alpha}\left[ L\circ (X\circ L)_- - (L\circ X)_-\circ
L\right]~.\(qAdler)\cr}$$
It is clear that $J^{(q)}(X)\in{\cal S}_q$, is linear in $X$ and has
thus has the general form $J^{(q)}(X) = \sum_{i,j=1}^\infty
(J_{ij}^{(q)}\cdot x_j)\xi^{q-i}$, for some differential operators
$J_{ij}^{(q)}$ defining, via \(Poissonbrackets), a $\W$-algebra
generated by the $\{u_i\}$.  The explicit expression for the
$J_{ij}^{(q)}$ can be found in \[WKPq], from where one can read that
this is a nonlinear centrally extended (for generic $q$) $\W$-algebra
of the $\Winfty$-type.  Moreover, for $q=1$ it agrees with $\W_{\rm
KP}$---hence it is a one-parameter deformation of it.  Therefore
$\Wkpq$ is a one-parameter family of nonlinear $\W$-algebras.

We can obtain a one-parameter deformation of $\widehat{\W}_\infty$
by reduction.  For $q\neq 0$, the constraint $u_1=0$ is formally
second-class and the Dirac brackets (denoted $\Omega_{ij}^{(q)}$)
define a $\W$-algebra on the remaining generators $\{u_{i>1}\}$; the
generator of spin 2 obeying a Virasoro algebra
$$\Omega_{22}^{(q)} = {\alpha\over12} q(q^2-1) \d^3 + u_2\d + \d
u_2~.\(Virasoro)$$
For $q=0$, $u_1$ decouples from the algebra and the remaining
generators obey a centerless nonlinear $\W$-algebra whose linear terms
are simply those of $\Winfty$.  We denote by
$\widehat{\W}^{(q)}_\infty$ the $\W$-algebra defined by
$\Omega_{ij}^{(q\neq 0)}$ or by $J_{ij}^{(0)}$ when $q=0$.

\section{The Structure of $\Wkpq$ for $q\in\nats$}

We already saw that $\W^{(1)}_{\rm KP} \isom \W_{\rm KP}$
and also that $\widehat{\W}^{(1)}_\infty \isom \widehat{\W}_\infty$.
We now investigate the structure of the algebras for $q=N$ a positive
integer $\geq 2$.  In this case it makes sense to impose the
constraint $L_-=0$.  It follows from \(qAdler) that $L_-=0 \implies
J^{(N)}(X)_-=0$ for all $X$, whence the $\{u_i\}_{i>N}$
generate an ideal $I_N^-$ of $\W^{(N)}_{\rm KP}$.  Since
$\W$-algebras are nonlinear, this does not mean that these generators
close among themselves: in the nonlinear terms they can appear
multiplied by arbitrary generators and a closer look at the explicit
form of the algebra shows that in fact they do.  A closer look also
reveals that there are no central terms in this ideal, whence the
linear and the quadratic terms must obey the Jacobi equations
separately---since they can both be obtained unperturbed as different
contractions of the ideal.  In particular, the linear terms define a
Lie algebra isomorphic to $\W_{\infty - N}$ without central
extension.  In \[Didier], Depireux conjectured the existence of a
nonlinear centerless deformation of $\W_{\infty-N}$ which he
calls $\widehat{\W}_{\infty-N}$.  It may be that redefining the
generators $u_i\mapsto\utilde_i= u_i + p_i(u_{j<i})$, the
$\W$-subalgebra of $\W^{(N)}_{\rm KP}$ generated by
$\{\utilde_{i>N}\}$ could be equivalent to $\widehat{\W}_{\infty-N}$.
It is however still an open problem to prove that the algebras
conjectured in \[Didier] actually exist.

Since $I_N^-$ is an ideal of $\W^{(N)}_{\rm KP}$, the quotient
is an algebra.  This is the $\W$-algebra generated by the first $N$
generators or, equivalently, by $L$ a differential symbol.  In other
words, this is precisely the Gel'fand--Dickey algebra $\GD_N$.  The
constraint $u_1=0$ is again formally second-class and upon reduction,
we recover $\W_N$.

We could equivalently first reduce to $\widehat{\W}_\infty^{(N)}$.
The same arguments as before imply that $\{u_{i>N}\}$ generate a
centerless ideal $\widehat{I}_N^-$ which again can be shown not to
close because of the nonlinear terms.  However the linear terms are
unchanged from those of $I_N^-$ and hence $\widehat{I}_N^-$ also
contracts to $\W_{\infty - N}^{c=0}$.

For $q\in\nats$ it also makes sense to deform the generalized Adler
map to obtain the analog of the first Gel'fand-- Dickey bracket.  This
bracket gives rise to a $\W$-algebra which can therefore be obtained
as a contraction of $\W_{\rm KP}^{(N)}$ as follows.  Let us shift $L$
by a constant $L\mapsto L + \lambda$ and call the resulting Adler map
$J^{(N)}_\lambda$.  From \(qAdler) it follows that
$$J^{(N)}_\lambda(X) = J^{(N)}(X) + \lambda
J^{(N)}_\infty(X)~,\(deformedAdler)$$
where $J_\infty^{(N)}$ is given by
$$ J_\infty^{(N)}(X) = \comm{L_+}{X_-}_+ -
\comm{L_-}{X_+}_-~.\(GDoneN)$$
This means that the deformed Poisson bracket contains a piece linear
in $\lambda$.  Since this shift corresponds to the change of variables
$u_N \mapsto u_N + \lambda$ and the Jacobi identity holds for
arbitrary $\{u_j\}$, it means that for all $\lambda$ the new bracket
satisfies the Jacobi identity.  This being a quadratic identity, it
means that it will contain pieces of orders 0, 1, and 2 in $\lambda$
which vanish separately.  In particular, the terms quadratic in
$\lambda$ are the Jacobi identities for the bracket defined by
$J^{(N)}_\infty$.  It follows from \(GDoneN) that the resulting
$\W$-algebra breaks up as a direct sum of commuting subalgebras
generated by the coefficients of the differential and integral parts
of $L$, respectively.  The $\W$-algebra generated by the coefficients
of the differential part is simply the first Gel'fand--Dickey bracket
$\GD^{(1)}_N$, whereas the one generated by the coefficients of the
integral part is precisely $\W_{1+\infty}$ without central
extension.  For all $\lambda$ the $\W$-algebra defined by
$J^{(N)}_\lambda$ is isomorphic to $\W_{\rm KP}^{(N)}$ and,
since $J^{(N)}_\infty = \lim_{\lambda\to\infty} \lambda^{-1}
J^{(N)}_\lambda$, we deduce that $\GD^{(1)}_N \times\W_{1+\infty}$ is a
contraction of $\W_{\rm KP}^{(N)}$.

For $N=1$ this result is simply the statement that the first
hamiltonian structure for the KP hierarchy is isomorphic to a
centerless $\W_{1+\infty}$ algebra.  For $N>1$, $\W_{\rm
KP}^{(N)}$ agrees with the family of hamiltonian structures defined by
Radul in \[Radul] (see also \[Das]) and their contraction are simply
the Watanabe structures studied in \[Watanabe].

We can summarize graphically our results for this section in the
following commutative diagram of $\W$-algebras\fnote{Strictly speaking
$I_N^-$ and $\widehat{I}_N^-$ are not $\W$-algebras but we include
them nevertheless for completeness.}:
$$\setbox1=\hbox{\ninerm ~contraction~}
\commdiag{\GD_N^{(1)} & \lflecha{contraction} & \GD_N&
\rflecha{reduction}&\W_N \cr
\mapup{\hbox{\ninerm truncation}}&&\mapup{\hbox{\ninerm reduction}}
&&\mapup{\hbox{\ninerm reduction}}\cr
\GD^{(1)}_N \times \W_{1+\infty}^{c=0}&\lflecha{contraction}
&\W^{(N)}_{\rm KP} & \rflecha{reduction} & \widehat{\W}_\infty^{(N)}\cr
\mapdown{\hbox{\ninerm truncation}}&&\mapdown{\hbox{\ninerm
truncation}} &&\mapdown{\hbox{\ninerm truncation}}\cr
\W^{c=0}_{1+\infty}&\lflecha{contraction}&I_N^-
&\rflecha{reduction}&\widehat{I}_N^-\cr
\mapdown{\hbox{\ninerm truncation}}&&\mapdown{\hbox{\ninerm
contraction}} &&\mapdown{\hbox{\ninerm contraction}}\cr
\W^{c=0}_{\infty-N}&\longequal&\W^{c=0}_{\infty-N}
&\longequal&\W^{c=0}_{\infty-N}\cr}$$
where the horizontal reductions all correspond to setting $u_1=0$ and
the horizontal contractions correspond to the deformation of the Adler
map induced by the shift in $u_N$.

\section{Some Contractions of $\Wkpq$}

The algebras $\W_{1+\infty}$, $\Winfty$, and its truncations,
obtained by contracting $\W_{\rm KP}^{(N)}$ are all centerless.
In order to obtain algebras with central extensions it is necessary to
perform the contraction at the same time as the limit $q\to N$ is
approached.  For example, if we take the limit $\alpha\to\infty$ and
$q\to 0$ in such a way that $q\alpha = c$ remains constant, an easy
calculation reveals the following expression for the limiting Poisson
structure
$$\lim_{q\to 0\atop \alpha q = c} J^{(q)}(\xi^q\circ X)\xi^{-q} =
-\comm{c\log\xi + L_-}{X_+}_-~,\(wopinftyc)$$
where $L$ is evaluated at $q=0$ in the right-hand-side of the
equation.  The $c$-independent term agrees with the similar term in
\(GDoneN) and as discussed in the previous section gives rise to a
centerless $\W_{1+\infty}$.  The $c$-dependent piece, however,
is nothing but the outer derivation on the algebra of
pseudodifferential symbols giving rise to the Khesin--Kravchenko
2-cocycle $c_{\lower3pt\hbox{\sevenrm KK}}(P,Q) = \Tr
\comm{\log\xi}{P}\circ Q$.  As shown in \[BKK] this gives precisely
the central terms in $\W_{1+\infty}$.  We let $\W_{1+\infty}^c$ denote
the central extension of $\W_{1+\infty}$ given by the
Khesin--Kravchenko cocycle.

Suppose we now take $q\to 1$ in such a way that $\alpha (q-1) = c$
remains constant.  Inspection reveals that $J_{11}^{(q)}$ diverges,
whence it is necessary to impose the constraint $u_1=0$, reducing to
$\widehat{\W}_\infty^{(q)}$.  An easy calculation reveals that the
limiting Poisson structure is exactly the one given above but where
$u_1$ does not appear.  Therefore it corresponds to the truncation of
$\W^c_{1+\infty}$ obtained by taking $\{u_{i>1}\}$ or, in other
words, $\Winfty^c$.

This fact generalizes.  If we take the limit $q\to N$ and
$\alpha\to\infty$ such that $\alpha (q-N) =c$, we find that the
central terms in $\Wkpq$ for $i,j\leq N$ all diverge in the limit.  We
must therefore reduce the algebra by setting them to zero.  The
resulting algebra--denoted tentatively\fnote{Our choice of notation
notwithstanding, we don't imply any relation between these nonlocal
algebras and the ones conjectured in \[Didier] besides the fact that
they are both deformations of $\W_{\infty-N}$.} by
$\widehat{\W}^{(q)}_{\infty - N}$) is nonlocal for all values of
$q$.  However in the limit, the nonlocal (as well as the nonlinear)
terms all vanish and we are left with the following expression for the
Poisson structure (here $\varepsilon = q-N$):
$$\lim_{\varepsilon\to 0\atop \alpha\varepsilon=c}
J^{(q)}(\xi^\varepsilon\circ X)\xi^{-\varepsilon} =
- \left( \comm{c\log\xi + (L\xi^{-N})_-}{(\xi^N \circ X)_+}\xi^N
\right)_-~,\(winftymnc)$$
where $L$ in the right hand-side is evaluated at $q=N$.  This Poisson
structure is identical to \(wopinftyc) except that only the
$\{u_{i>N}\}$ occur.  In other words, it is the subalgebra of
$\W^c_{1+\infty}$ generated by the $\{u_{i>N}\}$---namely,
$\W^c_{\infty - N}$.

Imposing the constraint $u_1=0$ on $\W^{c\neq 0}_{1+\infty}$ we
obtain a new nonlinear algebra we call $\Wnew$.  It follows that
$\Wnew$ is the contraction of $\widehat{\W}_\infty^{(q)}$ as $q\to
0$ and $\alpha q = c$.  One can show that this is a genuinely
nonlinear algebra in that there exists no polynomial field
redefinition which linearizes it.  Moreover it can be shown by
inspection of the first few Poisson brackets not to be equivalent to
$\widehat{\W}_\infty^{(q)}$ for any $q$.

We can again summarize these results diagrammatically:
$$\setbox1=\hbox{\ninerm ~contraction~$\scriptstyle q\to N$~}
\commdiag{\W_{\rm KP}^{(q)}&\longequal&\W_{\rm KP}^{(q)} &
\rflecha{contraction $\scriptstyle q\to 0$}&\W_{1+\infty}^c\cr
\mapdown{\hbox{\ninerm reduction}}&& \mapdown{\hbox{\ninerm
reduction}}&&\mapdown{\hbox{\ninerm reduction}}\cr
\widehat{\W}_{\infty-N}^{(q)}&\lflecha{reduction}&
\widehat{\W}_\infty^{(q)} &\rflecha{contraction $\scriptstyle q\to
0$}&\Wnew\cr
\mapdown{\hbox{\ninerm contraction}\atop {q\to N}}&&
\mapdown{\hbox{\ninerm contraction}\atop {q\to 1}}&&\cr
\W_{\infty-N}^c&\lflecha{truncation}&\W_\infty^c&&\cr}$$
where the horizontal arrows labelled reductions have the same meaning
as in the previous diagram.

\section{A Classical $\Winfty$ Tour}

As shown in \[Class], one can obtain classical limits of these Poisson
structure by taking the commutative limit of the algebra of
pseudodifferential symbols.  Rescaling $\xi$, we can introduce in
\(symbolcomp) a formal parameter $\hbar$ as follows:
$$P\circ Q= \sum_{k\geq 0} {\hbar^k\over k!}
\kpder{k}{P}{\xi} \kpder{k}{Q}{z}~.\(hbarcomp)$$
The classical limit of any structure is obtained by introducing the
parameter $\hbar$ via \(hbarcomp) and keeping only the lowest term in
its $\hbar$ expansion.  For example, the classical limit of $\circ$ is
simply the commutative multiplication of symbols.  Analogously, the
classical limit of the commutator is given by\fnote{We use $\dlb{}{}$
to the denote the Poisson bracket to avoid confusion with the Poisson
bracket defining the $\W$-algebras.}
$$\dlb{P}{Q} = \lim_{\hbar\to 0} \hbar^{-1} \comm{P}{Q} =
\pder{P}{\xi} \pder{Q}{z} -  \pder{P}{z}\pder{Q}{\xi}~,\(poisson)$$
which we recognize as the Poisson bracket relative to a
two-dimensional Darboux chart $z,\xi$.  The classical limit of the
generalized Adler map \(qAdler) is thus given by
$$J^{(q)}_{c\ell}(X) = \dlb{L}{X}_+ L - \dlb{L}{(XL)_+} =
\dlb{L}{(XL)_-} - \dlb{L}{X}_-L ~.\(Jcl)$$
Notice that these consist of the terms in $J^{(q)}_{ij}$ with exactly
one derivative.

It was shown in \[Class] that the dependence in $q$ in the classical
limit of $\W_{\rm KP}^{(q=N)}$ is fictitious and can be
eliminated by a polynomial redefinition of variables.  In \[WKPq] it
was shown that this result persists for $q\neq 0$.  In particular,
this shows that the classical limit of $\W_{\rm KP}^{(q)}$
is $\w_{\rm KP}$ for all $q\neq 0$ where this algebra is defined in
\[Class].  Analogously, and after the $u_1=0$ reduction, the
classical limit of $\widehat{\W}_\infty^{(q)}$ is independent of
$q$ and yields the reduction of $\w_{\rm KP}$ denoted \[WuYuKP]
$\widehat{\w}_\infty$.

Further contracting these algebras eliminates the nonlinear terms and
we recover $\w_{1+\infty}$ and $\w_\infty$---the classical limits of
$\W_{1+\infty}$ and $\Winfty$, respectively.  It is important to
remark that $\w_{1+\infty}$ appears without the central term in the
Poisson bracket of $u_1$.  In order to recover this central term we
have to perform the contraction more carefully.  We reintroduce the
parameter $q$ in $\w_{\rm KP}$, which in the classical limit is
inessential, and we take $q\to 0$ with $\alpha\to\infty$ in such a way
that their product remains constant.  In this limit we obtain
$\w^c_{1+\infty}$.

In $\w_{1+\infty}^c$ we can impose the constraint $u_1=0$ which is
formally second-class for $c\neq 0$.  The resulting algebra is the
classical limit $\w_\infty^\#$ of $\Wnew$.  The nonlinearity still
persists, but it can be shown that this algebra is not equivalent to
$\widehat{\w}_\infty$, as its spectrum would suggest.

The following commutative diagram of $\W$-algebras summarizes some of
these correspondences.
$$\setbox1=\hbox{\ninerm ~contraction~}
\commdiag{\W_{1+\infty}^c&\lflecha{contraction}&\Wkpq&
\rflecha{reduction}&\widehat{\W}_\infty^{(q)}&\rflecha{contraction}
&\Winfty^c\cr
\mapdown{\hbox{\ninerm classical}\atop\hbox{\ninerm limit}}&&
\mapdown{\hbox{\ninerm classical}\atop\hbox{\ninerm limit}}&&
\mapdown{\hbox{\ninerm classical}\atop\hbox{\ninerm limit}}&&
\mapdown{\hbox{\ninerm classical}\atop\hbox{\ninerm limit}}\cr
\w_{1+\infty}^c&\lflecha{contraction}&\w_{\rm KP}&
\rflecha{reduction}&\widehat{\w}_\infty&\rflecha{contraction}
&\winfty\cr
\Big\Vert&&\mapdown{\hbox{\ninerm reduction}}&&
\mapdown{\hbox{\ninerm reduction}}&&\Big\Vert\cr
\w_{1+\infty}^c&\lflecha{$\scriptstyle N\to\infty$}&\gd_N&
\rflecha{reduction}&\w_N&\rflecha{$\scriptstyle
N\to\infty$}&\winfty\cr}$$

\section{The Spectrum of $\Winfty$-type Algebras}

We end the paper with some comments on the spectrum of $\Winfty$-type
algebras---that is, how the generators transform under the action of
$\diffs1$.  In the introduction, the definition of what we mean by
$\Winfty$-type algebras was purposefully vague.  In the literature,
the accepted working definition is that of $\W$-algebra with one
generator for each spin $s\geq 2$ (and possibly also $s=1$).  This
usually connotes the existence of a Virasoro (or at least a $\diffs1$)
subalgebra and a choice of generators which transform tensorially
under the action of $\diffs1$ induced by this subalgebra.  In
$\W_n$---the prototypical example of a $\W$-algebra---this is indeed
the case, as it is also in its limit algebra $\winfty$.  In both
cases, the generator of lowest spin $u_2$ satisfies a Virasoro (for
$\W_n$) or $\diffs1$ (for $\winfty$) algebra and one can define an
action of $\diffs1$ as follows $\delta_\varepsilon u_j =
J_{j,2}\cdot\varepsilon$.  Under this action, the transformation laws
of the $\W_n$ generators contain terms with more than one derivative
acting on $\varepsilon$---whence the generators are not tensorial.
Nevertheless, as shown in \[dFIZ], there is a differential algebra
automorphism $u_j\mapsto w_j$ such that for $j>2$, $w_j$ is a tensor of
spin $j$.  The proof of this statement exploited the
$\diffs1$-covariance of the Lax operator (the differential operator to
which $L$ gets sent by \(symtops)) as well as the fact that $u_2$
transforms as a projective connection and can thus be put to zero in a
projective chart.  For $\winfty$---as for any classical limit of a
$\W$-algebra---the generators are automatically tensorial since the
classical limit of $J_{ij}$ are differential operators of order 1 and
therefore in $\delta_\epsilon u_j$ there cannot appear terms with more
than one derivative acting on $\varepsilon$.

It is tacitly assumed in the literature that this paradisiacal
situation persists for $\Winfty$-type algebras---albeit at the price
of introducing nonlinearities (see footnote in \[PRS]).  To the best
of our knowledge this has not been proven in any $\Winfty$-type
algebra (except for the classical limits in which there in nothing to
prove) and, in fact, it is not even true for $\widehat{\W}_\infty$.
It is conceivable, however, that if the two properties exploited in
\[dFIZ] for the $\W_n$ case still hold, one could prove the similar
result: the main difficulty being now the fact that we are not dealing
with differential operators, but with pseudodifferential ones instead.
As shown, for example, in \[dFIZ], the form of the Adler map and the
fact that the action of $\diffs1$ is hamiltonian relative to it
implies the covariance of the Lax operator.  It is therefore natural
to conjecture that, for nonvanishing Virasoro central charge, a
primary basis exists.

For simplicity, we shall restrict ourselves to discussing those
$\Winfty$-type algebras without the spin $1$ generator---namely
$\widehat{\W}_\infty^{(q)}$, $\Wnew$ and $\Winfty^c$.  Since $\Wnew$,
$\Winfty^c$, both arise as contractions of $\widehat{\W}_\infty^{(q)}$
for special values of the parameter, one can argue that to prove
existence of a primary basis for any of these algebras it is enough to
find a primary basis for $\widehat{\W}_\infty^{(q)}$ for generic $q$.
Treating $q$ as a formal parameter one finds by inspection of the
transformation laws under $\diffs1$, that the first few fields are
indeed redefinable into primaries.  These redefinitions, however,
involve adding to the fields differential polynomials whose
coefficients are rational functions of $q$---the dependence on $q$ of
the denominator occurring only through the Virasoro central charge in
\(Virasoro).  As a result, for those values of $q$ for which the
central charge vanishes, namely $q=0,\pm 1$, the redefinition becomes
singular and no longer legal.  This implies that $\widehat{\W}_\infty$
(and also $\widehat{\W}^{(0)}_\infty$ and
$\widehat{\W}_\infty^{(-1)}$) do not have a primary basis.  For the
contractions, however, things are different.  In the contracting
limits $q\to 0$ and $q\to 1$, the value of the central charge remains
finite and thus in the limit the redefinitions remain nonsingular.
Therefore a primary basis for $\widehat{\W}_\infty^{(q)}$ would induce
a primary basis on such a contraction.  For $q=N$, it follows that
both $\{u_{j\leq N}\}$ and $\{u_{j>N}\}$ generate $\diffs1$
submodules.  It follows from \[dFIZ] that the $\{u_{3\leq j\leq N}\}$
can be redefined into primaries.  Moreover $u_{N+1}$ is always a
primary and we have been able to redefine the first few higher $u_j$'s
into primaries; although we have not yet elaborated a general proof.
It is not clear to what extent covariance of the Lax operator plays a
crucial role since, as exemplified by the case of
$\widehat{\W}_\infty$, when one deals with pseudodifferential
operators, there are more ways to make a covariant operator than out
of tensors and covariant derivatives.

\ack

Sonia Stanciu helped us with some of the calculations and disproved
many an unenlightened conjecture.  We gladly take this oportunity to
express our gratitude to her.  Two of the authors (JMF and ER,
respectively) also take great pleasure in expressing their respective
thanks to the Instituut voor Theoretische Fysica of the Universiteit
Leuven and to the Departamento de F{\'\i}sica de Part{\'\i}culas
Elementales of the Universidad de Santiago de Compostela for their
hospitality and support during this collaboration.
\refsout
\bye